\documentclass{ws-ijmpaMY}
\usepackage{graphicx}

\usepackage{hyperref}    

\begin{document}

\markboth{A. A. Grib, Yu. V. Pavlov, V.D. Vertogradov}
{Geodesics with negative energy in the ergosphere of rotating black holes}

\catchline{}{}{}{}{}

\title{\large Geodesics with negative energy in the ergosphere\\
of rotating black holes}

\author{\normalsize A. A. Grib${}^{1,2,3,\,}$\footnote{E-mail:\,
andrei\_grib@mail.ru}, \ \ \ \ \
Yu. V. Pavlov${}^{1,2,4,\,}$\footnote{E-mail:\, yuri.pavlov@mail.ru}, \ \ \ \
V. D. Vertogradov${}^{3,\,}$\footnote{E-mail:\, askoli01@yandex.ru}}

\address{\vspace{4pt}
${}^{1}$A.\,Friedmann Laboratory for Theoretical Physics,
St.\,Petersburg, Russia\\[4pt]
${}^{2}$Copernicus Center for Interdisciplinary Studies,
Krak\'{o}w, Poland, EU\\[4pt]
${}^3$Theoretical Physics and Astronomy Department, The Herzen  University,\\
48 Moika, St.\,Petersburg, 191186, Russia\\[4pt]
${}^4$Institute of Problems in Mechanical Engineering, RAS,\\
61 Bolshoy, V.O., St.\,Petersburg, 199178, Russia}

\maketitle

\pub{Received 3 January 2014}{Revised 16 March 2014}
\vspace{-14pt}
\pub{Accepted 12 May 2014}{Published 12 June 2014}

\begin{abstract}
    It is shown that the geodesics with negative energy for rotating
black holes cannot originate or terminate inside the ergosphere.
    Their length is always finite and this leads to conclusion that they
must originate and terminate inside the gravitational radius of the ergosphere.

\end{abstract}

\ccode{PACS 04.70.-s, 04.70.Bw, 97.60.Lf}

\section{Introduction}
\label{secIntr}

     In this paper we analyse geodesics for particles with negative energies
for rotating black holes.
    Kerr's metric (see Ref.~\refcite{Kerr63}) predicts differently from
the Schwarzschild metric  existence of the special region outside the horizon
of the black hole called ergosphere.
    In the egosphere elementary particles must rotate together with the black
hole.
    The new feature of ergosphere is the existence of geodesics with negative
relative to infinity energy.
    Existence of such geodesics leads to the possibility of extraction of
energy from the rotating black holes due to the
Penrose process (see Ref.~\refcite{Penrose69})
    so one can call these geodesics the Penrose geodesics.
    However in spite of the more than 40 years passing after the discovery of
the Penrose effect there is still no information about the full picture of the
Penrose geodesics especially about their origin.
    Here we shall investigate the problem of properties of such geodesics.
    Particle can arrive on such trajectory in the result of collisions or
decays in the ergosphere.
    But the world line of the geodesic in geodesically complete space-time
must originate or terminate either in singularity or in
the infinity.\cite{HawkingEllis}
    Note that here one means infinity in space-time so that it can be infinity
in time for finite value of the space distances.
    The problem is to find where originate and terminate Penrose geodesics and
this is the subject of the paper.
    It will be shown that the length of the Penrose geodesics is always finite
inside the ergosphere so that the only possibility of their origination and
termination is outside of the ergosphere.
    But the geodesics for particles with negative energy don't exist in the
external space out of ergosphere so that one comes to the conclusion that they
originate and terminate inside the gravitational radius of the black hole.
    The fact that they terminate inside the horizon is well
known.\cite{Contopoulos84}
    One can ask why one cannot use the time reversal argument to conclude
that negative energy geodesics also originate inside the horizon?
    But Kerr's metric is not time reversal invariant due to the nondiagonal
term.
    So in general one can expect existence of geodesics in ergosphere with
negative energy originating in the infinite past going at some time inside
the horizon.
    In the paper we prove that this is not the case.

    The system of units $G=c=1$ is used in the paper.

\section{General formulas for the energy of particles
close to the black hole}
\label{sec2}

    The Kerr's metric of the rotating black hole in
Boyer--Lindquist coordinates has the form:\cite{BoyerLindquist67}
    \begin{eqnarray}
d s^2 &=& d t^2 -
\frac{2 M r}{\rho^2} \, ( d t - a \sin^2 \! \theta\, d \varphi )^2
\nonumber \\
&&-\, \rho^2 \left( \frac{d r^2}{\Delta} + d \theta^2 \right)
- (r^2 + a^2) \sin^2 \! \theta\, d \varphi^2,
\label{Kerr}
\end{eqnarray}
    where
    \begin{equation} \label{Delta}
\rho^2 = r^2 + a^2 \cos^2 \! \theta, \ \ \ \ \
\Delta = r^2 - 2 M r + a^2,
\end{equation}
    $M$ is the mass of the black hole, $ aM $ its angular momentum.
    The rotation axis direction corresponds to $\theta =0$, i.e. $a \ge 0$.
    The event horizon of the Kerr's black hole corresponds to
    \begin{equation}
r = r_H \equiv M + \sqrt{M^2 - a^2} .
\label{Hor}
\end{equation}
    For Cauchy horizon one has
    \begin{equation}
r = r_C \equiv M - \sqrt{M^2 - a^2} .
\label{HorC}
\end{equation}
    The surface of the static limit is defined by
    \begin{equation}
r = r_1(\theta) \equiv M + \sqrt{M^2 - a^2 \cos^2 \theta} .
\label{Lst}
\end{equation}
    In case $ a \le M $ the region of space-time between the static limit
and event horizon is called ergosphere.

    For geodesics in Kerr's metric~(\ref{Kerr}) one obtains
(see Ref.~\refcite{Chandrasekhar}, Sec.~62 or
Ref.~\refcite{NovikovFrolov}, Sec.~3.4.1)
    \begin{equation} \label{geodKerr1}
\rho^2 \frac{d t}{d \lambda } = -a \left( a E \sin^2 \! \theta - J \right)
+ \frac{r^2 + a^2}{\Delta}\, P,
\end{equation}
    \begin{equation}
\rho^2 \frac{d \varphi}{d \lambda } =
- \left( a E - \frac{J}{\sin^2 \! \theta} \right) + \frac{a P}{\Delta} ,
\label{geodKerr2}
\end{equation}
    \begin{equation} \label{geodKerr3}
\rho^2 \frac{d r}{d \lambda} = \sigma_r \sqrt{R}, \ \ \ \
\rho^2 \frac{d \theta}{d \lambda} = \sigma_\theta \sqrt{\Theta},
\end{equation}
    \begin{equation} \label{geodP}
P = \left( r^2 + a^2 \right) E - a J,
\end{equation}
    \begin{equation} \label{geodR}
R = P^2 - \Delta [ m^2 r^2 + (J- a E)^2 + Q],
\end{equation}
    \begin{equation} \label{geodTh}
\Theta = Q - \cos^2 \! \theta \left[ a^2 ( m^2 - E^2) +
\frac{J^2}{\sin^2 \! \theta} \right].
\end{equation}
    Here $E$ is conserved energy (relative to infinity)
of the probe particle,
$J$ is conserved angular momentum projection on the rotation axis
of the black hole,
$m$ is the rest mass of the probe particle, for particles with nonzero
rest mass $\lambda = \tau /m $,
where $\tau$ is the proper time for massive particle,
$Q$ is the Carter's constant.
    The constants $\sigma_{r}, \sigma_{\theta}$ in formulas~(\ref{geodKerr3})
are equal to $\pm 1$ and are defined by the direction
of particle movement in coordinates $r$, $\theta$.
    For massless particles one must take $m = 0$
in~(\ref{geodR}) and (\ref{geodTh}).

    The permitted region for particle movement is defined by conditions
    \begin{equation} \label{ThB0}
R \ge 0, \ \ \ \ \ \Theta \ge 0, \ \ \ \ \
\frac{d t}{d \lambda} \ge 0 .
\end{equation}
    The last inequality forbids movement ``back in time''.\cite{Wald}
    Let us find limitations for the particle angular momentum from the
conditions~(\ref{ThB0}) at the point $(r, \theta)$,
taking the fixed values of $\Theta$.\cite{GribPavlov2013}

    Outside the ergosphere $ r^2 -2 r M +a^2 \cos^2 \! \theta >0 $ one obtains
    \begin{equation} \label{EvErg}
E \ge \frac{1}{\rho^2} \sqrt{(m^2 \rho^2 + \Theta)
(r^2 -2 r M +a^2 \cos^2 \! \theta)},
\end{equation}
    \begin{equation} \label{JvErg}
J \in \left[ J_{-}, \ J_{+} \right],
\end{equation}
    \begin{eqnarray}
J_{\pm} = \frac{\sin \theta}{r^2 -2 r M +a^2 \cos^2 \! \theta}
\biggl[ - 2 r M a E \sin \theta \ \ \ \
\nonumber \\
\pm \sqrt{ \Delta \left( \rho^4 E^2 \!-\! (m^2 \rho^2 \!+\! \Theta)
(r^2 \!-\! 2 r M \!+\! a^2 \cos^2 \! \theta) \right)} \biggr] . \ \ \
\label{Jpm}
\end{eqnarray}

    On the boundary of ergosphere
    \begin{equation} \label{rEgErg}
r = r_1(\theta) \ \ \ \Rightarrow \ \ \ E \ge 0,
\end{equation}
    \begin{equation} \label{JgErg}
J \le E \left[ \frac{M r_1(\theta) }{a} + a \sin^2 \! \theta \left(\!
1 - \frac{m^2}{2 E^2} - \frac{\Theta}{4 M r_1(\theta) E^2} \!\right)
\! \right]\!.
\end{equation}

    Inside ergosphere
    \begin{equation} \label{lHmdd}
r_H < r < r_1(\theta) \ \ \ \Rightarrow \ \ \
(r^2 -2 r M +a^2 \cos^2 \! \theta) <0 ,
\end{equation}
    \begin{eqnarray}
J \le J_{-}(r,\theta) = \frac{- \sin \theta}{
r^2 \!-\! 2 r M \!+\! a^2 \cos^2 \! \theta} \biggl[ 2 r M a E \sin \theta \ \
\nonumber \\
- \sqrt{ \Delta \left( \rho^4 E^2 \!-\! (m^2 \rho^2 + \Theta)
(r^2 \!-\! 2 r M \!+\! a^2 \cos^2 \! \theta) \right)} \biggr]. \ \
\label{JmErg}
\end{eqnarray}
    So it is only inside the ergosphere that the energy $E$ of the particle
relative to infinity can be negative.
    From~(\ref{JmErg}) one can see that for negative energy~$E$ of the particle
in ergosphere its angular momentum projection on the rotation axis of the
black hole must be also negative.

    In the limit $r \to r_H$ from~(\ref{JmErg}) (for $\theta \ne 0, \pi$)
one obtains
    \begin{equation} \label{JgEH}
J \le J_H = \frac{ 2 M r_H E}{a} .
\end{equation}
    So $J_H$ is the maximal value of the angular momentum of the particle with
the energy~$E$ close to the gravitational radius.

\section{Properties of movement of particles with negative energy
in ergosphere}
\label{secErgo}

    For negative values of the energy $E$ the function $ J_{-}(r,\theta) $
is decreasing with growing $r$ in ergosphere, so that
    \begin{equation} \label{JgEHf}
\theta \ne 0 , \pi , \ \ \ r \to r_1(\theta) \ \ \Rightarrow \ \
J_{-}(r,\theta) \to - \infty .
\end{equation}
    So in order to come to the upper frontier of the ergosphere particle
with negative energy one must have infinitely large in absolute value negative
angular momentum.

    From~(\ref{geodKerr3}), (\ref{geodTh}), (\ref{JmErg}) one can see that the
Carter's constant $Q \ge 0$ for particles with negative energy in ergosphere.
   One can have $Q=0$ for $E \le0$ only in case of movement in
equatorial plane.
    Note, that for $E > 0$ the nonequatorial trajectory exists
with $Q=0$ (see Ref.~\refcite{Chandrasekhar}).

    Let us investigate other properties of movement for the negative energy
particles in ergosphere.
    Define the effective potential by the formula
    \begin{equation} \label{Leff}
V_{\rm eff} = - \frac{R}{2 \rho^4}.
\end{equation}
    Then due to~(\ref{geodKerr3})--(\ref{geodR}),
    \begin{equation} \label{LeffUR}
\frac{1}{2} \left( \frac{d r}{d \lambda} \right)^{\!2} + V_{\rm eff}=0
\end{equation}
    and so
    \begin{equation} \label{LeffUR2}
\frac{d^2 r}{d \lambda^2} = - \frac{d V_{\rm eff}}{d r} .
\end{equation}
    The necessary condition of existence of orbits with constant $r$ is
    \begin{equation} \label{LeffCucl}
V_{\rm eff}=0, \ \ \ \ \frac{d V_{\rm eff}}{d r} =0\,.
\end{equation}

    Let us show that for particles with nonzero and zero masses (photons) with
negative relative to infinity energy there are no orbits inside ergosphere
with constant $r$ or with $r$ changing for all geodesic inside
the interval $r_1 \ge r \ge r_H$.
    The proof for equatorial geodesics was given by us
earlier.\cite{GribPavlov2013a}
    Here we give the proof for the general case.
    To prove our statement it is sufficient to show that for
$dt/d \lambda >0$ and
    \begin{equation} \label{LeffdVefg0}
E<0, \ \ \ r > r_H, \ \ \ V_{\rm eff}(r)=0 \ \ \Rightarrow
\ \ V^{\, \prime}_{\rm eff}(r) > 0 .
\end{equation}

    For $V_{\rm eff}(r)=0$ the derivative of the effective potential
can be written as
    \begin{equation} \label{Leffder}
V_{\rm eff}^{\, \prime}(r) = \frac{1}{2 \rho^4}
\left( 2 (r -M) \frac{P^2}{\Delta} + 2 m^2 r \Delta - 4 r E P \right).
\end{equation}
    So in order to prove our statement it is sufficient to prove that
for the negative energy of the particle in ergosphere one has $P >0$.

    From the condition of movement ``forward in time'' and~(\ref{geodKerr1})
one obtains
    \begin{equation} \label{geodKett}
P \ge  \frac{ a \left( a E \sin^2 \! \theta - J \right) \Delta}
{r^2 + a^2}.
\end{equation}
    For particles in ergosphere with $E<0$ from~(\ref{lHmdd}), (\ref{JmErg})
one has
    \begin{equation}
a E \sin^2 \! \theta - J \ge a E \sin^2 \! \theta  - J_{-} \ge
\frac{a E \sin^2 \! \theta }{ r^2 -2 r M +a^2 \cos^2 \! \theta} >0.
\label{LPerg}
\end{equation}
    That is why $P>0$ and our statement is proved.

    So there are no circular orbits for Penrose trajectories
in Kerr's black holes.
    The permitted zone for such particles in ergosphere can have only
upper boundary.

    Note that one gets absence of orbits with constant $r=r_H$ on the horizon
of the nonextremal black holes $a<M$ from the fact that for $V_{\rm eff}=0$
and $\Theta \ge 0$ one  has $J=2 r_H /M$ and
    \begin{eqnarray}
V^{\, \prime}_{\rm eff}(r_H)=
\frac{(r_H - M) (m^2 r_H^2 + (J-a E)^2 + Q}{\rho^4} \ge
\nonumber \\
\frac{(r_H-M)}{\rho^4 a^2} ( m^2 \rho^2 a^2 +
(r_H^4 - a^4 \cos^2 \! \theta ) E^2 + 4 r^2_H M^2 \cot^2 \! \theta ) > 0 .
\label{VderH}
\end{eqnarray}
    For extremal black holes $a=M$ one can see from~(\ref{VderH})
$V^{\, \prime}_{\rm eff}(r_H)=0$ for $\theta= \pi/2$.
    However circular orbits for $E\ne 0$ are also absent in this
case as it is shown in Ref.~\refcite{HaradaKimura10}.

\section{The time of movement of particles with negative
energy in the ergosphere}
\label{secErttau}

    Let us analyze the problem of the time of movement for particles with
negative energy in ergosphere.
    As it was shown in the previous section the geodesic with negative energy
in ergosphere begins from $r=r_H$, then achieves the upper point of the
trajectory $r_b$ and after it falls to horizon.
    So the proper time interval of movement of the particle along all geodesic
in ergosphere is defined by the integral
    \begin{equation} \label{VsIntdlbO}
\Delta \lambda = 2 \int \limits_{r_H}^{r_b} \frac{ d r }{|d r /d \lambda |}
= 2 \int \limits_{r_H}^{r_b} \frac{d r}{\sqrt{- 2 V_{\rm eff}(r)}} .
\end{equation}
    The factor 2 before the integral is due to taking into account the fact
that the proper time of movement along geodesic up from some value of
the radial coordinate $r$ is equal to the time of falling down to the same
value of $r$.

    In the vicinity of the upper point $r_b$
on the trajectory of the particle with negative energy one has
from (\ref{LeffUR}) and $V_{\rm eff}(r_b)=0$ that
    \begin{equation} \label{VsdrdlrH}
\left| \frac{d r}{d \lambda} \right| = \sqrt{- 2 V_{\rm eff}} \approx
\sqrt{2 (r_b - r) V^{\, \prime}_{\rm eff}(r_b)}.
\end{equation}
    As it was shown in the preceding part for the boundary point of the
permitted zone $V^{\, \prime}_{\rm eff}(r_b)>0$, so the integral
    \begin{equation} \label{VsIntdlb}
\int \frac{ d r }{|d r /d \lambda |} \sim
\int \frac{d r}{\sqrt{2 (r_b - r) V^{\, \prime}_{\rm eff}(r_b)}}
\end{equation}
    is convergent and the proper time of the lifting to the upper point
(falling from the upper point) of the trajectory in the vicinity of this
point is finite.

    Due to the fact that permitted zones for particles with negative energies
in ergosphere can have only upper boundary there are no zeros
for $dr / d \lambda$ in the other points of the trajectory.
    So the integral~(\ref{VsIntdlbO}) is convergent and the proper time of
movement along geodesic in the ergosphere for the particle with the negative
energy is finite.

    Note that the coordinate time of movement to the horizon is infinite.
    For equatorial movement the evaluations of the divergence of the
integral for coordinate time were given in Ref.~\refcite{GribPavlov2013a}.

    Finiteness of the proper time for movement of particles with negative
energy in ergosphere of the black hole leads to the problem of the
origination and termination of such trajectories.
    As we said in the Introduction these lines cannot arrive to ergosphere
from the region outside of the ergosphere.
    So they must originate and terminate inside the gravitational radius.
    This means that they originate as ``white hole'' geodesics originating
inside the horizon.

    Note that similar situation takes place for some geodesics of particles
with positive energy in Schwarzschild metric
(see the text book Ref.~\refcite{LL_II}).
    The geodesic completeness leads to the necessity of taking into account
``white hole'' geodesics originating in the past singularity of the eternal
black hole for radial geodesics with specific energy $E/m<1$
arising from the region inside the gravitational radius.
    However for Penrose geodesics we show that all such geodesics
in ergosphere of the Kerr's black hole have such behaviour.

    Inside the event horizon up to Cauchy horizon
from~(\ref{geodKerr3}), (\ref{geodR}), (\ref{geodTh}), (\ref{Leff})
one has $V_{\rm eff} <0$ for any falling particles.
    So any particle intersecting the event horizon must achieve the
Cauchy horizon.
    After going through the Cauchy horizon the particle can achieve
singularity.
    The necessary condition for this is that Carter constant $Q \le 0$
(see~(\ref{geodKerr3})--(\ref{geodTh})).
    For particles with negative energy in ergosphere this is true only for
movement in equatorial plane, i.e. $Q=0$
according to Carter's theorem.\cite{Carter68}
    From~(\ref{geodKerr3})--(\ref{ThB0}), (\ref{JgEH}) one can see that all
light like particles (photons) with negative energy falling in equatorial
plane from ergosphere achieve singularity.
    Massive particles also achieve singularity for example if $E=-m$
or the angular momentum is such that
    \begin{equation} \label{FalSin}
J \le J_H \biggl( 1 + \frac{a}{2M} \sqrt{1 + \frac{m^2}{E^2} } \, \biggr).
\end{equation}
    The proofs of all these results is the same for
$d r / d \lambda >0$ and $d r / d \lambda < 0$.
    That is why the same conditions are valid for ``white hole''geodesics
originating in Kerr's singularity, arriving to ergosphere and then going back
inside the gravitational radius.

    All particles with nonequatorial movement and $E \le 0$ don't achieve
singularity.
    Particles moving in equatorial plane also do not achieve singularity
if for example
    \begin{equation} \label{FalNS}
\frac{|E|}{m} \ll \frac{r_C}{M} , \ \ \frac{|J|}{m M} \ll \frac{r_C}{M} .
\end{equation}
    Then after achieving some minimal values of the radial coordinate
the particle can turn it's movement in the direction of larger $r$ and come
back to ergosphere along the white hole geodesics.

    As it is shown in Ref.~\refcite{GribPavlov2014} the energy in the centre
of mass frame of collisions of particles with negative energies coming from
the region inside the gravitational radius to ergosphere with ordinary
particles is growing to infinity on the horizon.
    The same phenomenon exists for the Schwarzschild black holes if particle
on the ``white hole geodesic'' collide with ordinary particles falling
on the horizon.
    The effect occurs due to the unlimited growth of the Lorentz factor of
the relative velocity of two particles in the centre of mass frame.

\section{Conclusion}
\label{Conclusion}

1) Geodesics with negative energy originate inside the gravitational radius
of the rotating black hole and are ``white hole'' geodesics!
    This is similar to the case of Schwarzschild eternal black hole for which
it was shown in Ref.~\refcite{GribPavlov2010NE} that negative energy
trajectories arise in the white hole past singularity.
    However differently from the eternal Schwarzschild black hole when there
are two different space like singular surfaces --- the black hole and white
hole --- here we have one Kerr's time like singular surface on which some
geodesics originate and some terminate.

2)
One can get information about the interior of the gravitational radius if
some particles move along these geodesics!
    So there is no cosmic censorship for such rotating black holes if one
understands cosmic censorship as impossibility to get information from
the region inside the gravitational radius.
    The cosmonaut can get direct information about the interior of the
gravitational radius only inside the ergosphere.
However if one considers interaction of negative energy particles radiated by
the ``black-white'' hole to ergosphere with ordinary positive energy particles
escaping the ergosphere this information can be obtained by any external observer.
    The electromagnetic interaction of the negative energy photons with usual
matter can lead to some new physical process of ``annihilation'' of this
matter inside ergosphere.
    The energy in the centre of mass frame of the collision of ordinary
positive energy particle with the particle with negative energy is growing
without limit on the horizon.

3)
The mass of the ``black-white'' hole can grow due to the radiation of negative
energy particles to ergosphere where these particles interact with
positive energy particles.

    Surely all results in this paper correspond to exact Kerr's solution.
    For physically realized rotating black holes as the result of the star
collapse due to nonstability of interior solution inside the gravitational
radius especially inside the Cauchy horizon
(see Ref.~\refcite{GribPavlov2008UFN}) the results can be different and
the special research is needed.

\section*{Acknowledgments}

A.A.G. and Yu.V.P. were supported by the grant from
The John Templeton Foundation.


\end{document}